\documentclass[preprint,superscriptaddress,pre,floats,aps,amsmath,amssymb,12pt]{revtex4-1}

\usepackage[utf8]{inputenc} 
\usepackage[T1]{fontenc}    
\usepackage{booktabs}       
\usepackage{amsfonts}       
\usepackage{nicefrac}       
\usepackage{microtype}      
\usepackage{graphicx}			 
\usepackage[usenames,dvipsnames]{xcolor}
\usepackage{multirow}
\usepackage{balance}
\usepackage{times,mathptmx}
\usepackage{lastpage}
\usepackage{fnpos}

\begin{document}

\title{Experimental determination of the gadolinium L subshells fluorescence yields and Coster-Kronig transition probabilities}

\author{Yves Kayser}
\email[E-mail: ]{yves.kayser@ptb.de}
\affiliation{Physikalisch-Technische Bundesanstalt, Abbestrasse 2-12, 10587 Berlin, Germany.}
\author{Malte Wansleben}
\altaffiliation[Present address: Helmut Fischer GmbH, R\"{o}ntgen- und Neutronenoptiken, Rudower Chaussee 29/31, 12489 Berlin, Germany. E-mail: ]{malte.wansleben@helmut-fischer.com}
\affiliation{Physikalisch-Technische Bundesanstalt, Abbestrasse 2-12, 10587 Berlin, Germany.}
\author{Philipp H\"{o}nicke}
\author{Andr\'{e} W\"{a}hlisch}
\author{Burkhard Beckhoff}
\affiliation{Physikalisch-Technische Bundesanstalt, Abbestrasse 2-12, 10587 Berlin, Germany.}

\begin{abstract}
The L subshell Coster Kronig (CK) transition factors of Gd have been experimentally determined by means of two different experimental approaches and compared to available literature data. On the one hand reference-free X-ray fluorescence (XRF) analysis using an energy-dispersive detector was applied. This method permitted in addition to determine the L subshell fluorescence yields since the absolute incident and emitted X-ray photon flux can be determined. On the other hand a wavelength dispersive spectrometer based on a modified von Hamos geometry using a full cylinder highly annealed pyrolithic graphite crystal for an experimental validation of the CK transition factors in an independent experiment. The use of this high energy resolution spectrometer allowed thanks to the capability to better discriminate the different X-ray emission lines to validate the reference-free XRF results. For both experiments the use of calibrated instrumentation enabled the provision a reliable uncertainty budget on the results obtained. 
\end{abstract}

\keywords{atomic fundamental parameters \and rare earth metals \and gadolinium \and X-ray spectrometry \and von Hamos spectrometer}

\maketitle

\section{Introduction}
Atomic fundamental parameters (FPs) are the basis for the description of light-matter interaction. In the X-ray range the atomic FPs include mass attenuation coefficients, photionization cross-sections, fluorescence yields, Coster-Kronig (CK) factors as well as emission line energies, widths and relative transition probabilities. The atomic FPs are element and in general photon energy (sub)shell specific. The determination of the atomic FPs requires suitable and well-defined samples as well as dedicated experiments to be conducted for their accurate determination such as transmission or fluorescence experiments with (preferably calibrated) energy- or wavelength dispersive spectrometers. Nowadays, different technological developments allow for more accurate and reliable experiments for the determination of atomic FPs. These advances are in the field of sample preparation techniques, 
X-ray detection instrumentation, where improved sensitivity, signal-to-noise ratio, energy resolution, dynamic range and linear response behavior are offered, and the availability of tunable monochromatic X-ray radiation of high spectral purity. These improved capabilities can be expected to lead to more consistent data between different experiments barring experimental artefacts and inconsistent data treatment. Thus, an independent validation of experimental and theoretical results (or interpolation thereof) reported in databases such as \cite{Bearden,Deslattes,Krause1979,Elam1,xraylib,CAMPBELL2003} is required. With respect to the large uncertainties reported or interpolation of experimental and / or theoretical data, this endeavor can be even be considered as mandatory in order to have reliable and consistent data in future experiments. An improved and accurate understanding of X-ray interaction with matter will be beneficial for fundamental sciences, quantitative XRF analysis and calculations of atomic systems to test limits and needs of single-particle approaches and perturbative approaches. 

In this respect a substantial part of the research activities of Prof. Jean-Claude Dousse was directed to an accurate determination of different atomic FPs using reflection-type or transmission-type high energy-resolution crystal-based spectrometers \cite{Hoszowska96,Szlachetko13} and stand-alone laboratory based X-ray sources, synchrotron radiation or charged particles for the excitation of the X-ray emission processes to be studied. Examples of this work include the study of linewidths of X-ray emission processes \cite{Hoszowska1994}, radiative Auger transitions \cite{Herren96}, X-ray resonant Raman scattering \cite{Szlachetko2006}, CK factors \cite{Cao2009,Cao2010}, hypersatellite X-ray transitions \cite{Hoszowska2010,Maillard2018}, two-electron one-photon transitions \cite{Hoszoswka2011}, atomic level widths \cite{Fennane2013} and off-resonant X-ray spectroscopy \cite{Blachucki2014}. In this work, an independent investigation of the CK factors of the Gd L shell using two different detection schemes, a radiometrically calibrated silicon drift detector (SDD) and a full cylinder von Hamos spectrometer, is reported as a further illustration of the valuable contributions of wavelength dispersive spectrometry to an accurate determination of atomic FPs.


\section{Coster-Kronig transition factors}
The CK factors are included in the Sherman equation \cite{Sherman1955} which is the theoretical basis for quantitative X-ray fluorescence (XRF) since it describes the basics of the production and emission of XRF photons from thin one-elemental foils. Indeed, it can be considered as the chronological order of the physical processes occurring: from the number and energy of the monochromatic incident X-ray photons, the incidence angle on the sample surface, the thickness and density of the considered one-elemental foil and the XRF production factor $\sigma$ for a selected electronic subshell it is possible to calculate the number of produced XRF photons produced after an (usually dipole allowed) electronic transition 
to the subshell where the vacancy was created to minimize the total energy of the system (atom). When the solid angle of detection $\Omega/4\pi$ and the detection efficiency $\epsilon$ are known the quantitative amount of the isotropically emitted XRF photons which is detected can be calculated. The XRF production factor $\sigma$ is hereby defined as the product of the incident photon energy dependent photoionization cross section of inner shell electron, the fluorescence yield, which provides the probability for a radiative relaxation, the relative transition probability in case only fluorescence photons originating from the electronic transition between selected subshells from different electronic levels is of interest, and the CK factor(s) for all but the innermost subshell of an electronic level. The CK factor takes into account the relative transition probability for a non-radiative transition of the vacancy created from an inner to an outer subshell of an electronic level such that the principal quantum number of the vacancy does not change \cite{COSTER1935}. A CK transition must be energetically allowed in the sense that the energy difference between the subshells between which the vacancy is transferred is large enough to promote an outer electron to the continuum. This intrashell transition, which can also be described as an Auger transition within the same electronic level, leads hence to an increase in the number of vacancies created in the outer subshells before the electronic transition during which an Auger electron or XRF photon will be emitted takes place. Such multiply ionized atoms can be encountered as well in plasma physics and astrophysics. But this consideration is also important if the number of atoms contributing to the XRF signal is used for quantification purposes by using a fundamental parameter based approach \cite{deBoer89} or the reference-free X-ray spectrometry (XRS) method \cite{Beckhoff2008} but present as well a very sensitive probe to a correct modelling of initial- and final-state wavefunctions and electron binding energies \cite{Bambynek72,Chen81}. 

For the definition of the CK factors the description of Jitschin et al. is used \cite{W.Jitschin1985}. When considering the L subshells and using an incident photon energy above the L$_1$ ionization threshold, a possible transfer of vacancies from the L$_1$ and L$_2$ subshells to higher L subshells is possible. The fluorescence production factor for each L subshell is then written as follows:
\begin{align}
   \sigma_{\text{L}_3}(E_0) &= \omega_{\text{L}_3}\left[ \tau_{\text{L}_3}(E_0) + f_{2,3} \tau_{\text{L}_2}(E_0) + (f_{1,3}+f_{1,2}f_{2,3}) \tau_{\text{L}_1}(E_0)\right] \label{eq:sigma_L_3}\\
   \sigma_{\text{L}_2}(E_0) &= \omega_{\text{L}_2}\left[ \tau_{\text{L}_2}(E_0) + f_{1,2} \tau_{\text{L}_1}(E_0)\right] \label{eq:sigma_L_2}\\
   \sigma_{\text{L}_1}(E_0) &= \omega_{\text{L}_1}\left[ \tau_{\text{L}_1}(E_0) \right] \label{eq:sigma_L_1}
\end{align}
where $E_0$ is the incident photon energy, $\omega_{\text{L}_i}$ is the fluorescence yield, $\tau_{\text{L}_i}(E_0)$ the photoionization cross section of the subshell L$_i$ and $f_{i,j}$ the CK factor (with $i=1,2$ or $3$ and depending on the value of $i$, $j=2$ or $3$ ). For incident photon energies below the ionization threshold of subshell L$_i$, the corresponding CK factor $f_{i,j}$ is equal to zero. By using tunable monochromatic X-ray sources, readily available at synchrotron radiation facilities, it is thus possible to turn the CK transitions selectively on or off by a changing the energy of the incident photons \cite{W.Jitschin1985}, which enables determining individually the different CK factors while respecting the boundary conditions that $f_{i,j} \leq 1$ and $f_{1,3}+f_{1,2} \leq 1$. 

The CK factors were determined in the past by means of coincidence measurements, which are restricted to the $f_{2,3}$ factor, using radionuclides \cite{Bambynek72} or simultaneously photoinduced K and L XRF \cite{Santra04,Dunford06} or by means of a selective excitation by detecting Auger electrons or XRF photons using a tunable excitation source \cite{W.Jitschin1985,Sorensen91}. The former approach is suitable for obtaining results with low uncertainties but restricted in terms of elements for which the CK factors can be investigated, the latter approach is affected by larger errors which are mainly connected to the an incomplete knowledge on the photoionization cross-section of the different subshells \cite{Papp}. Still the access to all relevant subshell transition yields is granted. The trade-off between Auger and XRF resides in the complexity of the spectra generated and, hence, the availability of a priori information required for a proper separation of the contributions originating from the different L subshells. For XRF based works, mostly energy-dispersive detectors were used \cite{W.Jitschin1985,Stotzel1992,Barrea2004,M.Kolbe2012,M.Kolbe2015,Menesguen18,Menesguen20} but wavelength-dispersive spectrometers for measurements on selected X-ray emission lines were used as well \cite{Cao2009,Cao2010}.

\section{Experimental}
The reference-free XRF and X-ray emission spectroscopy (XES) experiments for the determination of the fundamental parameters of interest here were carried out at PTB’s four-crystal-monochromator (FCM) beamline \cite{Krumrey1998}. It provides monochromatized radiation between 1.75 keV and 10.5 keV by means of either four InSb(111) or Si(111) crystals. The contribution of higher harmonics in the spectral range applied is well below $10^{-5}$ \cite{Krumrey1998}. The incident photon energy was varied from 7.2$\,$keV (below the Gd-L$_3$ edge to exclude sample contamination) up to 9.0$\,$keV (above the Gd-L$_1$ edge) in steps of 50$\,$eV using the Si(111) crystals for monochromatizing the X-ray radiation originating from a bending magnet. The beam diameter in the focal plane of the beamline, where the measurements were performed, was about 300 $\mu m$.

A thin Gd coating with a thickness of 250$\,$nm on the top of a 500$\,$nm thick silicon nitride membrane acting as support material was used as sample. The Gd layer was coated with a thin 40$\,$nm thick Al layer in order to protect it from oxidation. The thickness selected for the Gd layer ensures a good transmittance in the photon energy regime of the Gd-L absorption edges and emission lines, i.e., the attenuation is of the order of 10$\%$ 
when considering the incidence and detection angles used. 
From the transmission experiments it was possible to determine sample specfic mass attenuation in order to not rely on theoretical data
. In addition, good measurement statistics for the fluorescence and emission experiments were granted, and secondary excitation effects and background contributions from the Al layer and silicon nitride membrane are also negligible. 

Furthermore, it should be avoided that the photoionization process involves electrons that are more tightly bound to the atom than those in the electronic level considered for the calculation of the XRF photon rate. Otherwise cascade effects (transition of the vacancy created from inner to outer electronic shells) and possible secondary fluorescence effects do not need to be considered. This condition is usually granted by the large energy difference between the ionization thresholds of the K shell and the different L subshells. A different factor, which can inherently not be avoided for all elements, is possible secondary fluorescence once incident photon energies above the L$_2$ or L$_1$ ionization threshold are used, resulting in the emission of XRF photons with sufficient energy to create a vacancy in the L$_3$ and / or the L$_2$ subshell. For Gd, this is the case for the L$\gamma$ emission lines for which the line energies and relative transition probabilities were established in recent works \cite{Wansleben, Menesguen}: the L$\gamma_2$ (L$_1$N$_2$) and L$\gamma_3$ (L$_1$N$_3$) emission lines resulting from a vacancy in the L$_1$ subshell have energies above the L$_2$ ionization threshold with a relative transition probability of 0.166(17) and 0.091(9), respectively and the the L$\gamma_1$ (L$_2$N$_4$) emission line resulting from a vacancy in the L$_2$ subshell has an energy above the L$_3$ ionization threshold with a relative transition probability of 0.154(15). However, due to the relatively low transition probabilities of the involved emission lines and the low sample thickness the expected secondary enhancement effect is only in the order of one permille.

\subsection{Transmission measurements}
\begin{figure}[!h]
  \centering
    \includegraphics[width=8cm]{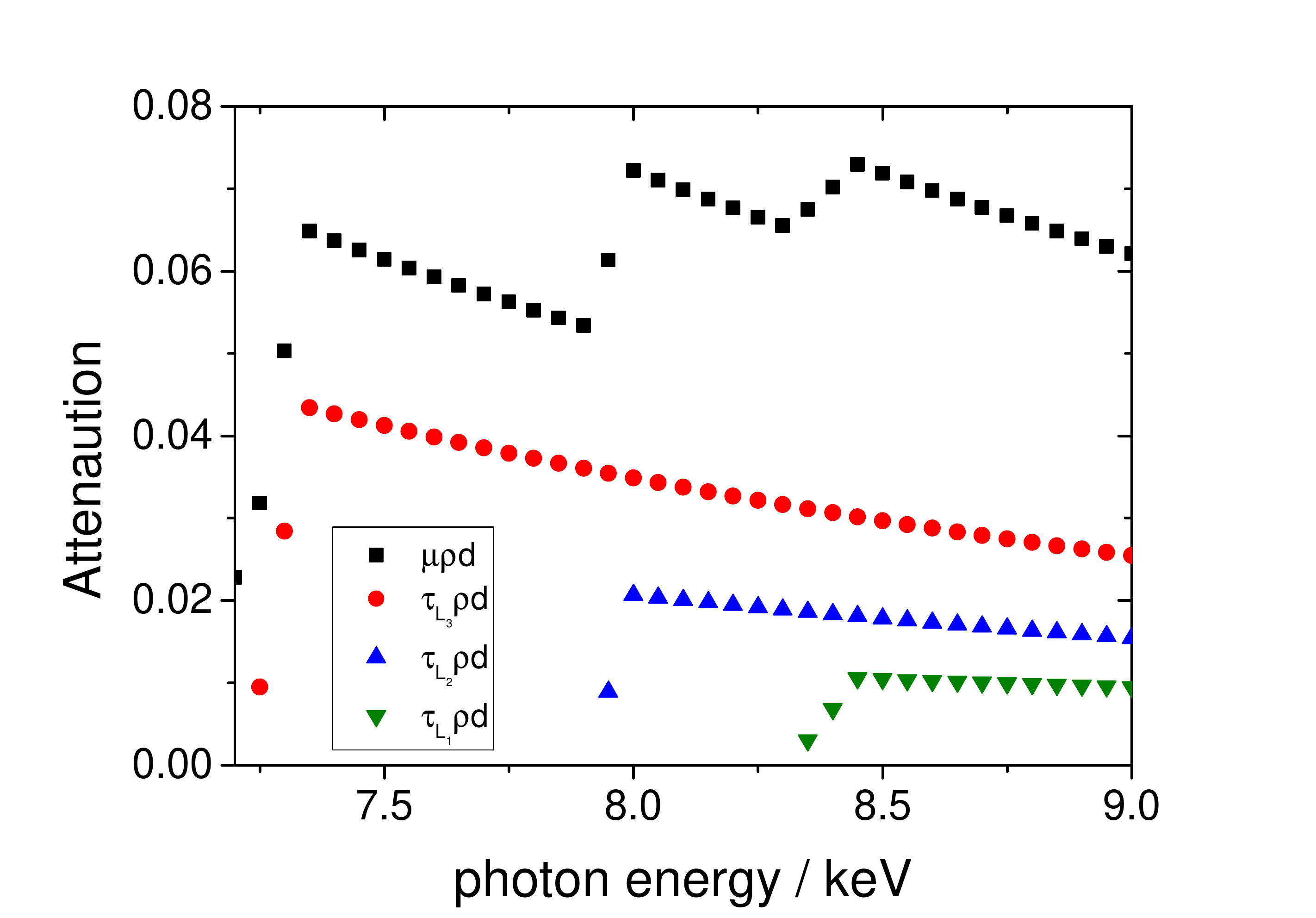}
  \caption{Incident photon energy dependence of the product of the mass attenuation, respectively photoionization cross section with the mass deposition as obtained for the 250$\,$nm thick Gd film used. In the selection of the film thickness care must be taken that the transmittance remains within a range of 20$\%$ to 80$\%$ for the sake of precision of the measurement.}
  \label{fig:cross-sections}
\end{figure}

The determination of the CK factors $f_{i,j}$ and the fluorescence yield $\omega_{\text{L}_i}$ depends on the absolute knowledge of the incident photon energy dependence of subshell photoionization cross-sections $\tau_{\text{L}_3}(E_0)$ and $\tau_{\text{L}_2}(E_0)$. To this respect transmission measurements with the sample were performed using a diode downstream of the sample location at different incident photon energies. To correct for the substrate and cap layer, a bare substrate with the same thin Al cap layer was measured as well. A subsequent subtraction of inelastic and elastic scattering cross-sections as reported by Ebel \cite{Elam1} and using a 5th order polynomial interpolation following the approach described in ref. \cite{Unterumsberger2018} allowed obtaining the product $\tau_{\text{L}_i}(E_0) \rho d$ (Fig. \ref{fig:cross-sections}). With this experiment no tabulated or theoretical values for the photon energy dependent mass attenuation coefficients at the relevant incident photon energies and exact information on the density and thickness of the foil are needed and thus unknown uncertainty contributions are avoided.
The relative error on uncertainty of the product $\tau_{\text{L}_i}(E_0) \rho d$ is estimated to be 3$\%$ for the L3 subshell and 5$\%$ for the L2 and L1 subshells.

\subsection{Reference-free XRF}
For the experiments, an in-house developed ultrahigh vacuum chamber dedicated to reference-free XRS was used \cite{J.Lubeck2013}. The incident photon flux was determined using a thin transmission diode upstream and a calibrated photodiode \cite{A.Gottwald2006} downstream of the sample position. The thin transmission diode allowed monitoring flux variations during each measurement, while a cross-referencing to the calibrated photodiode for each incident photon energy used allowed to determine the absolute photon flux for each measurement with a relative uncertainty of 1.5$\%$. A second, non-calibrated diode positioned downstream of the sample position was used for the transmission measurements.

\begin{figure}[!h]
  \centering
    \includegraphics[width=13cm]{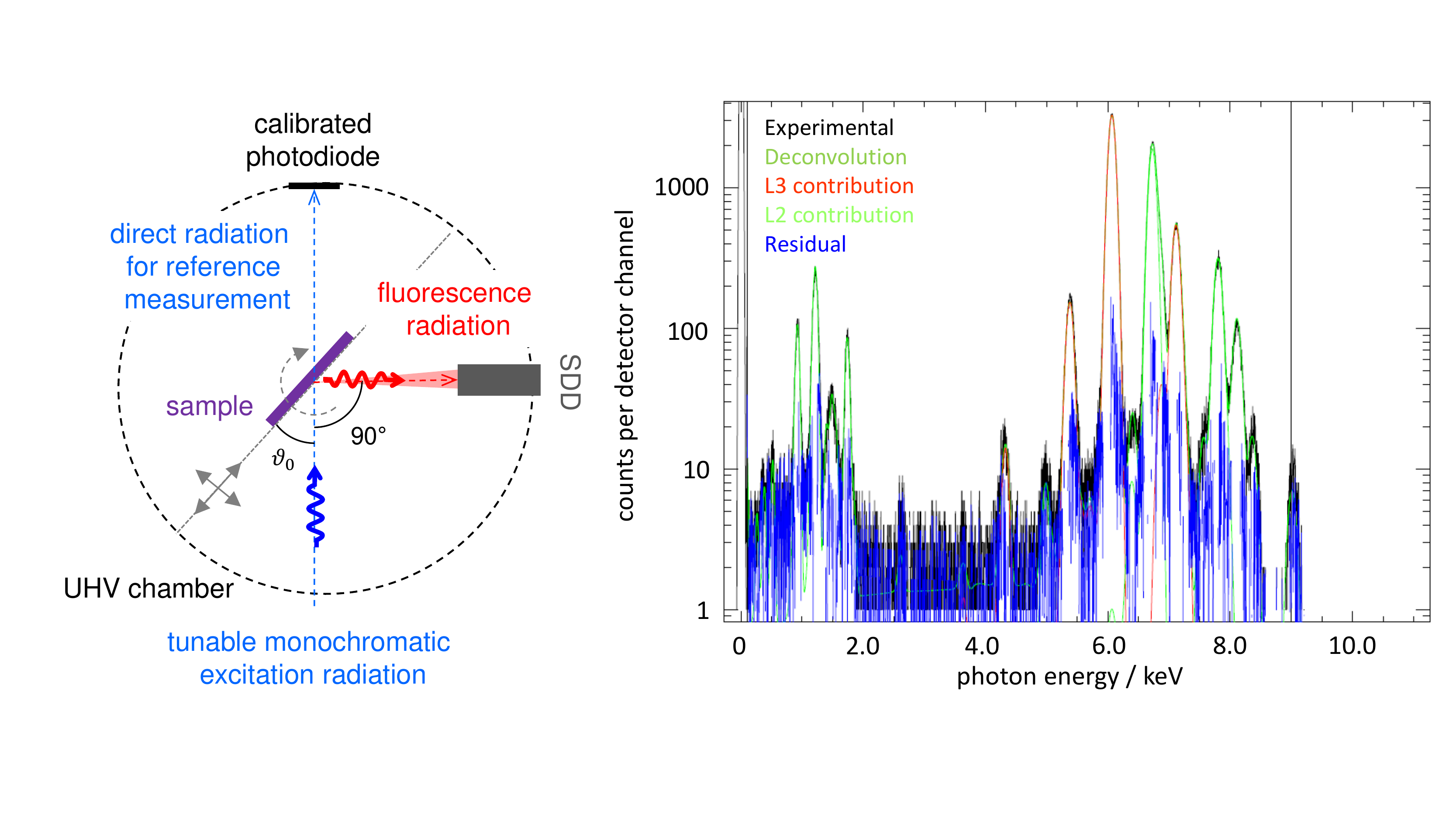}
  \caption{Exemplary Gd L XRF spectrum measured with a radiometrically calibrated SDD at an excitation energy higher than the L$_1$ subshell ionization threshold (9 keV). The knowledge of the detector's response function permits for a deconvolution of the XRF spectrum using only physically relevant contributions.}
  \label{fig:sdd_entfaltung}
\end{figure}

The sample was positioned with its surface plane intersecting the pivot point of the goniometer installed within the ultrahigh vacuum chamber and oriented such that the incidence angle of the monochromatized synchrotron radiation and the detection angle of the XRF radiation were each 45$^\circ$ (Fig. \ref{fig:sdd_entfaltung}). The experimental chamber is positioned such that the pivot point of the goniometer is located on the axis along which the synchrotron radiation propagates. The fluorescence radiation emitted by the sample is detected by means of a radiometrically calibrated silicon drift detector (SDD) \cite{Scholze2005,F.Scholze2009}, which is placed behind a well-characterized aperture in a well-defined distance from the sample. 
The diameter of the aperture was determined by an optical microscope, and defined together with the sample-detector distance the solid angle of detection with a relative uncertainty of 0.7$\%$. Since the detector is positioned within the polarization plane and perpendicular to the propagation direction of the incident synchrotron radiation, scattering contributions in the recorded spectra are minimized. The detector response functions as well as the detection efficiency of the SDD were known \cite{Scholze2005,F.Scholze2009}, the latter with a relative uncertainty of 1.5$\%$. The intensity of the fluorescence lines of interest were derived from the recorded spectra by a spectral deconvolution using physically motivated detector response functions. The natural Lorentzian broadening of the fluorescence lines could be neglected here due to the resolving power of the detector, which is about 2 orders of magnitude larger.

For the determination of the different CK factors corresponding to the electronic transitions between the different Gd L subshells a similar strategy as described in refs. \cite{M.Kolbe2012,M.Kolbe2015} was followed: XRF spectra were recorded with the SDD while varying the incident photon energies from beneath the L$_3$ ionization threshold to above the L$_1$ ionization threshold and were subsequently deconvolved and normalised to the incident photon flux, detector live time, solid angle of detection and energy-dependent detector efficiency to extract the absolute emission rate of XRF photons appertaining to an electronic transition to the L$_1$, L$_2$ and L$_3$ subshell respectively. For this purpose, the spectra recorded at incident photon energies between the L$_3$ and L$_2$ ionization thresholds were analyzed first since only fluorescence lines appertaining to electronic transitions to the L$_3$ subshell contribute to the recorded spectra. Thus, the relative transition probabilities of the different Gd L$_3$ emission lines that could be resolved by the SDD were accurately determined by means of a spectral deconvolution of each spectrum recorded when considering the contributions of of each individual fluorescence line and of relevant background contributions. By using the information from the spectral deconvolution for different incident photon energies the consistency of the established relative transition probabilities could be verified against possible incident photon energy dependent background or scattering contributions. The relative transition probabilites were then used as a fixed input parameter for incident photon energies above the L$_3$ ionization threshold: the emission lines appertaining to an electronic transition to the L$_3$ subshell were not allowed to vary individually in the spectral deconvolution, only the total net intensity of the ensemble of these emission lines was determined in the following after the spectral deconvolution and data normalization. The same procedure was applied to the XRF fluorescence lines appertaining to a transition to the L$_2$ subshell and L$_1$ subshell for incident photon energies between the L$_2$ and L$_1$ ionization thresholds .
This approach allowed reducing the number of free parameters in the deconvolution procedure such that its numerical stability was improved and the contribution of the XRF from each subshell to the measured spectrum could be established in a robust manner (Fig. \ref{fig:sdd_entfaltung}). Finally, the strategy described using sets with fixed relative intensities for the emission lines connected to an electronic transition to the different subshells allowed deriving the respective absolute detected count rates with uncertainties of 1.5 \% by the spectral deconvolution \cite{M.Kolbe2012}. Note that only the determination of the fluorescence yields necessitates the determination of absolute fluorescence emission rates for each subshell, the CK factors can be obtained from relative measurements as shown in the XES experiments. Moreover, and as discussed in ref. \cite{M.Kolbe2012}, it is assumed that the polarization of the incident radiation does not lead to anisotropic emission of the L shell emission lines beyond the uncertainties associated with the detection and deconvolution.

\subsection{X-ray emission spectroscopy}
In an independent experiment, which followed a similar concept, a wavelength dispersive spectrometer in a full-cylinder von Hamos geometry (as introduced in ref. \cite{Anklamm2014}) was made use of for the determination of the CK factors (but not the L subshell fluorescence yields). The spectrometer used \cite{Holfelder2022} is characterized by a medium to high energy resolution such that a good separation of the different Gd emission lines is ensured. In this respect the spectrometer was already applied to the determination of the energy and width of the Gd L emission lines as well as the relative intensity ratios of the L emission lines from the different L subshells \cite{Wansleben}. The spectrometer is attached to a measurement chamber similar to the one described in ref. \cite{J.Lubeck2013} and additionally equipped with an SDD. The sample orientation was similar as during the reference-free XRF experiments, i.e., the incidence angle on the sample surface was 45$^\circ$ and the dispersion axis of the spectrometer was perpendicular to the propagation direction and contained within the polarization plane of the incident synchrotron radiation. The dispersion axis is defined as the axis of the full cylinder highly annealed pyrolithic graphite (HAPG) crystal used. 
The SDD was positioned mid-way between the incidence direction of the synchrotron radiation and the detection direction of the von Hamos spectrometer.

\begin{figure}[!h]
  \centering
    \includegraphics[width=13cm]{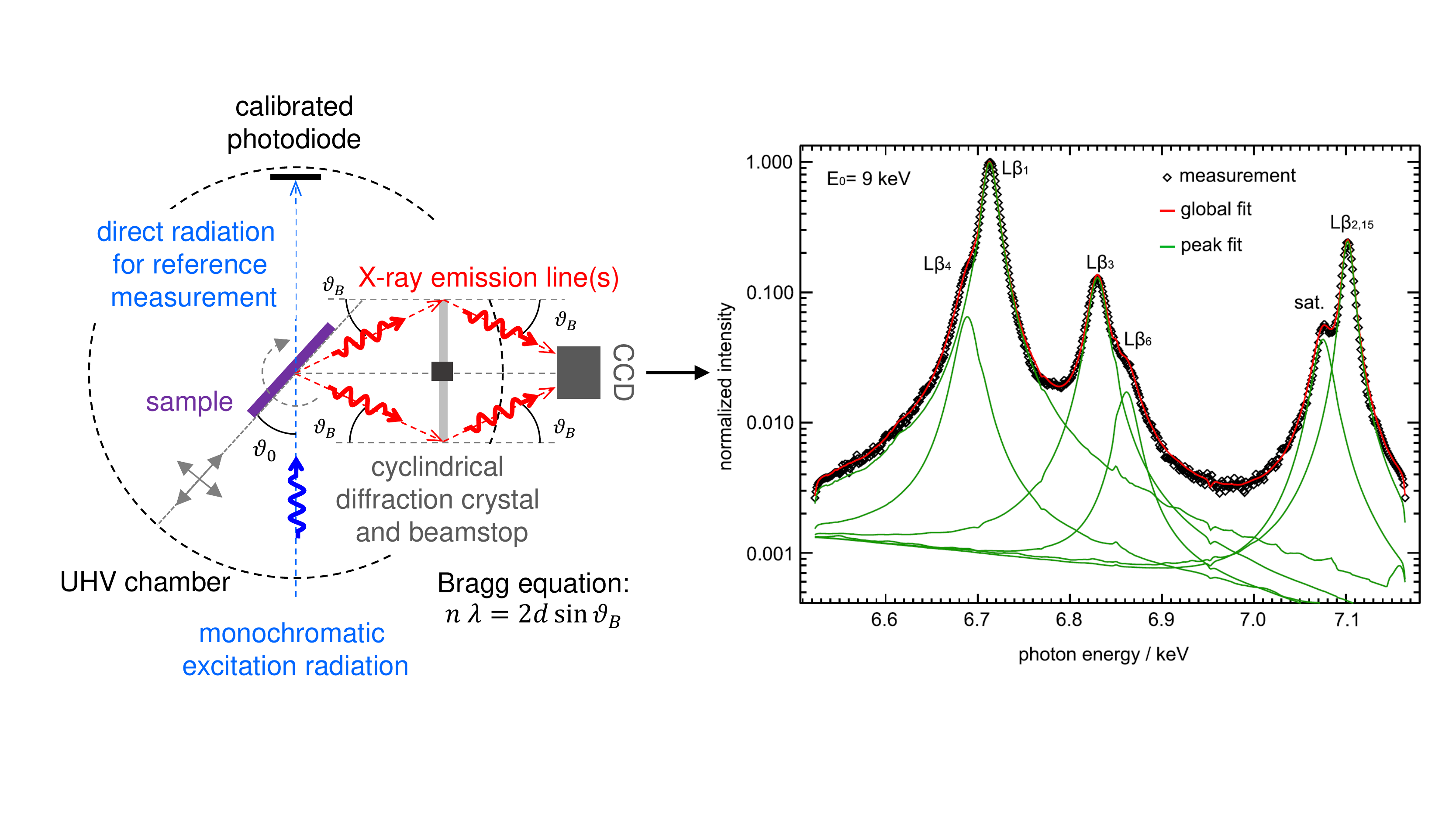}
  \caption{Exemplary Gd L$\beta$ emission spectrum measured with the von Hamos spectrometer at an excitation energy higher than the L$_1$ subshell ionization threshold (9 keV). The fit is a convolution of experimentally determined response functions and Lorentzians and a linear background.}
  \label{fig:vH_entfaltung}
\end{figure}

The spectrometer is operated in a slitless geometry, which makes the use of micro-focusing optics necessary, and equipped with two full-cylinder HAPG crystals with a bending radius of 50$\,$mm and a CCD camera as a detector \cite{Holfelder2022}. The incident monochromatized synchrotron radiation was therefore focused by means of a polycapillary half-lens resulting in a 60$\,\mu$m FWHM focus as determined with the knife-edge method \cite{R.Unterumsberger2012} at the sample position. 
In the presented experiment only one optic was used as this configuration provided a good enough resolving power of $E/\Delta E$ = 1000 to resolve the emission lines of interest. 
Moreover, due to the small radius of curvature of the crystals and the width of the crystals and size of the CCD chip, the advantage of a large spectral energy window of over 600$\,$eV at a single position of the HAPG crystal and the CCD camera can be exploited. The energy window mentioned is monitored in a scanning-free mode due to the cylindrical crystal shape and the position-sensitivity of the detector used.

The CK factors were determined by solely measuring a part of the the Gd L$\beta$ emission spectrum which includes emission lines originating from ionization of all three L subshells: L$_2$M$_4$ (L$\beta_1$), L$_3$N$_1$ (L$\beta_6$) and L$_3$N$_{4,5}$ (L$\beta_{2,15}$). The emission lines L$_1$M$_2$ (L$\beta_4$), L$_1$M$_3$ (L$\beta_3$) are also included in the recorded spectra but not required for the determination of the CK factors. In addition, a satellite emission on the low-energy side of the L$\beta_{2,15}$ is observed, which is due to exchange and spin-orbit interactions between the 4$d$ and 4$f$ electronic levels and discussed in a previous work \cite{Wansleben}. As in the reference-free XRF experiments, the different X-ray emission lines monitored enable the direct measurement of changes in emitted intensity of the emission lines associated with an electronic transition to the L$_3$ or L$_2$ subshell when increasing the incident photon energy above the L$_2$ and L$_1$ ionization thresholds. The deviations from the expected energy-dependent photoionization cross-sections of the L$_3$, respectively L$_2$ subshell are directly connected to the CK transitions. Since all the emission lines are monitored at once for each incident photon energy a more accurate normalization to the incident photon flux is enabled than compared to sequential measurements of the emission lines. 

\begin{figure}[!h]
  \centering
    \includegraphics[width=8cm]{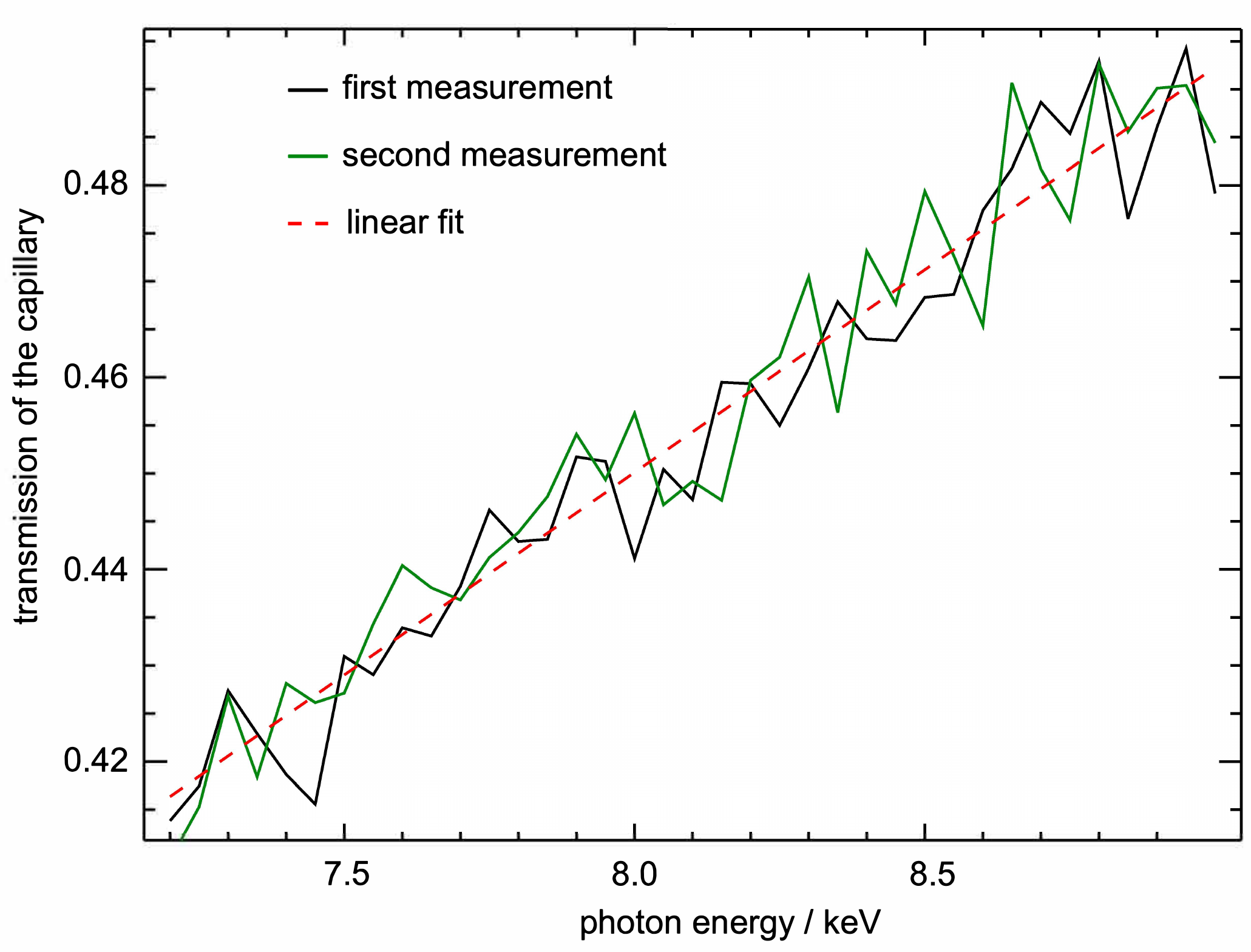}
  \caption{The derived transmission of the capillary as a function of all applied incident energies. }
  \label{fig:trans_cap}
\end{figure}

In contrast to the reference-free XRF measurements, the energy dependent transmission of the polycapillary optics needs to be taken into account when normalizing the data acquired at the different incident photon energies to the incident photon flux. Therefore, the intensity of the Ti K$\alpha$ emission lines from a metallic Ti foil was measured using the SDD with and without the polycapillary inserted in the incident beam path. This procedure is conducted for all incident photon energies before and after the acquisition of the Gd L$\beta$ emission spectrum with the von Hamos spectrometer in order to ascertain the stability and reproducibility of the alignment of the polycapillary optics. Fig. \ref{fig:trans_cap} shows the derived transmission of the capillary. This approach was found to be more stable and less affected by experimental noise than a comparable measurement with a diode, which needs to be positioned at an adequate distance of the capillary optics to avoid both depletion and illuminating an area larger than the diode area when focusing the X-ray beam. Finally, incident X-ray photon intensity variations during the XES measurements were monitored with the same transmission diode positioned upstream of both the sample and the polycapillary optics as in the reference-free XRF experiments. By referencing the transmission diode to an absolutely calibrated photodiode for each photon energy used, the absolute photon flux on the sample could be determined for each measurement. The relative uncertainty on the incident photon flux is about 3.1$\%$ and is determined by the counting statistics during the determination of the transmission of the polycapillary optics (relative contribution of 2.7$\%$) and the diode measurements (relative contribution of 1.5$\%$, as for the reference-free XRF measurements). Note, that an absolute determination of the incident photon flux is not strictly required for the determination of the CK factors. However, the use of radiometrically calibrated diode allows to alleviate possible incident photon energy dependent changes in sensitivity and thus for an improved accuracy of the results provided.

To extract the respective line intensities the measured Gd L$\beta$ spectra are deconvolved using the experimentally determined spectrometer's response function as well as Lorentzians that model the natural line shape of the individual emission lines. The response functions are measured prior to the Gd L$\beta$ emission by introducing a 4 $\mu$m-thick gold foil as scattering target and tuning the monochromator energy to the expected emission line energies of the Gd L$\beta$ spectrum. The choice of Au was motivated by the absence of X-ray emission lines in the energy range monitored and the large scattering cross-sections of high-Z elements in the X-ray energy range considered, which enabled the use of thin foils in view of optimizing the inelastically versus elastically scattered radiation. This measurement using elastically scattered X-ray photons is necessary as the response function slightly changes as a function of the emission energy \cite{Anklamm2014,Holfelder2022}. The respective calibration procedure is described in more details in ref. \cite{Wansleben}. In addition to the X-ray emission lines, a linear background is included in the deconvolution algorithm. An exemplary fit of the recorded spectra is displayed in Fig. \ref{fig:vH_entfaltung} showing the Gd L$\beta$ emission lines with the incident energy higher than the L$_1$ subshell threshold (9 keV). Using the residuals the relative uncertainty on the deconvolution of the L$\beta_{2,15}$ (including the satellite transition) and the L$\beta_{1}$ lines, which are the most intense lines in the recorded spectra, is estimated to be about 7$\%$ while the counting statistics is not a relevant contribution in the uncertainty budget.

\section{Results \& Discussion}
The CK factors were determined on the basis of the Sherman equation for quantification of the elemental mass
\begin{equation}
    \rho d=\frac{{N_{\text{L}_i}(E_0)}\sin(\theta) M(E_0,E_{\text{L}_i})}{I_0 \sigma_{\text{L}_i}(E_0) \Omega/(4\pi) \epsilon(E_{\text{L}_i}) }
    \label{eq:Sherman}
\end{equation}
where $\rho$ is the mass density of the sample, $d$ the sample thickness, $\theta$ the incidence angle of the synchrotron radiation on the sample, $M(E_0,E_{\text{L}_i})$ the correction factor for attenuation of incident and emitted X-rays within the sample (as defined in ref. \cite{M.Kolbe2015}, less than 10$\%$ for the energies considered in the present experiments), $N_{\text{L}_i}(E_0)$ the detected count rate of the fluorescence (e.g., for the L$_3$ subshell the ensemble of emission lines appertaining to an  electronic transition to the L$_3$ subshell) using an incident photon flux $I_0$, and $\epsilon(E_{\text{L}_i})$ the detection efficiency of the detector used for the emission lines considered. In the case of XES measurements where only selected emission lines are monitored $N_{\text{L}_i}(E_0)$ would need to be corrected for the relative transition probability.

\begin{figure}[!h]
  \centering
    \includegraphics[width=16cm]{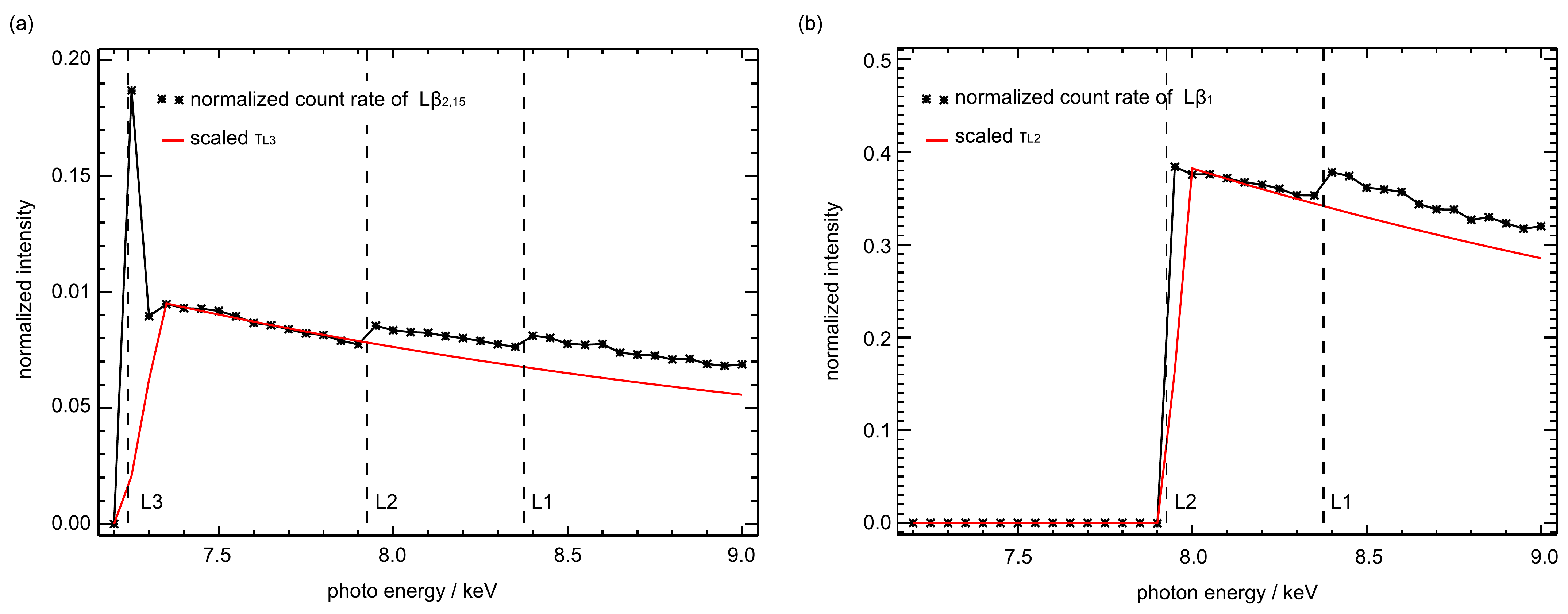}
  \caption{Comparison of the subshell photoionisation cross section $\tau$ and the normalized count rate of a respective emission line: (a) L$\beta_{2,15}$ (L$_3$N$_{4,5}$ transition) for the L$_3$ subshell and (b) L$\beta_1$ (L$_2$M$_4$ transition) for the L$_2$ subshell. The jumps in the count rate when tuning the photon energy over the next subshell threshold indicates the CK transitions. }
  \label{fig:Gd_L_tau_floureszenz}
\end{figure}

Since the mass deposition $\rho d$ is invariant, the CK factors can be determined by using different incident photon energies above and below the respective L subshell ionization thresholds and inserting equations \ref{eq:sigma_L_3} and \ref{eq:sigma_L_2} into \ref{eq:Sherman} assuming that $\sigma_{\text{L}_i} M(E_0,E_{\text{L}_i}) /\omega_{\text{L}_i}$ is proportional to the measured count rates normalized to the incident flux $I_{\text{L}_i}(E_0)=N_{\text{L}_i}(E_0)/I_0$. In Fig. \ref{fig:Gd_L_tau_floureszenz}, the intensity jump at the ionization thresholds for the emission lines monitored is due to vacancies created by a CK transition in addition to the photoionization process. Thus, taking into account the incident photon energy dependency of the individual subshell photoionization cross sections, a pair of equations can be established that can be solved for the CK factors,
\begin{align}
f_{23} &= \frac{\tau_{\text{L}_3}(E_B)}{\tau_{\text{L}_2}(E_B)} \left [ \frac{\tau_{\text{L}_3} (E_A)}{I_{\text{L}_3}(E_A) } \frac{I_{\text{L}_3}(E_B)}{\tau_{\text{L}_3}(E_B)} - 1 \right ]
\label{eq:f23} \\
f_{12} &= \frac{\tau_{\text{L}_B}(E_C)}{\tau_{\text{L}_C}(E_C)} \left [ \frac{\tau_{\text{L}_2} (E_B)}{I_{\text{L}_2}(E_B) } \frac{I_{\text{L}_2}(E_C)}{\tau_{\text{L}_2}(E_C)} - 1 \right ]
\label{eq:f12} \\
f_{13} &= \frac{\tau_{\text{L}_3}(E_C)}{\tau_{\text{L}_1}(E_C)} \left [ \frac{\tau_{\text{L}_3} (E_A)}{I_{\text{L}_3}(E_A) } \frac{I_{\text{L}_3}(E_C)}{\tau_{\text{L}_3}(E_C)} - \left (1 + f_{23} \frac{\tau_{\text{L}_2}(E_C)}{\tau_{\text{L}_3}(E_C)} \right ) \right ] - f_{12}f_{23},
\label{eq:f13}
\end{align}
when using incident photon energies such that $E_{\text{L}_3} < E_A < E_{\text{L}_2} < E_B < E_{\text{L}_1} < E_C$. The uncertainties then depend mainly on the relative contributions of the photoionization cross-section $\tau_{\text{L}_i}$ and the deconvolution result $I_{\text{L}_i}$ or the ensemble of emission lines appertaining to an electronic transtion to subshell L$_i$ for the reference-free XRF experiment, respectively for the XES results by using $I(\text{L}\beta_{2,15})$ for $I_{\text{L}_3}$ and $I(\text{L}\beta_{1})$ for $I_{\text{L}_2}$. Additionally, for each CK factor multiple different excitation energies were used that fulfill the above written condition. Hence, for each CK factor multiple results are obtained which allowed for a consistency check of the CK factors determined, for example whether the incident photon energy dependent trend for the subshell photoionization cross-sections $\tau_{\text{L}_i}(E_0)$ is correctly extrapolated. The individual results for each CK factor scattered on a range of 6$\%$ to 8$\%$ which is well within the uncertainty for, both reference-free XRF and XES. The final result reported for each CK factor is an average over these multiple results. This procedure leads to more robust results and provides a consistency check of the experimental results, which is important for assessing whether an incident photon energy dependence attributable to the subshell photoionization cross-sections used may be present. The results obtained for the XES and reference-free XRF measurements are summarized and compared to literature values in Table \ref{tab:CKs}. 


\begin{table}[htbp]
  \centering
  \caption{L subshell Coster Kronig factors}
    \begin{tabular}{|c|c|c|c|}
    \hline
          & $f_{2,3}$ & $f_{1,2}$ & $f_{1,3}$ \\
\hline
    Present work (RF-XRF) & 0.235(59) & 0.169(96) & 0.197(121) \\
\hline
    Present work (XES) & 0.206(60) & 0.148(43) & 0.193(147) \\
\hline
    M\'{e}nesguen et al. (2020) \cite{Menesguen20} & 0.095(20)  & 0.053(23) &  0.27(4) \\
\hline
    Puri et al. (1993) \cite{S.Puri1993} & 0.16  & 0.216 &   0.334 \\
 \hline
    X-RayLib (2012) \cite{xraylib} & 0.149  & 0.19  & 0.279 \\
\hline
    Krause (1979) \cite{Krause1979} & 0.147 &  0.19 &  0.3  \\
\hline
    Papp et al. (1998) \cite{Papp_1998}  &              & 0.166(20)  & 0.287(14) \\
\hline
    Douglas (1972) \cite{Douglas_1972}& 0.223(11) &             &         \\
\hline
    Gnade et al. (1981) \cite{Gnade_1981} & 0.157(12) &          &  \\
\hline
    \end{tabular}%
  \label{tab:CKs}%
\end{table}%

First it can be noticed that the CK factors extracted from the reference-free XRF and the XES experiments agree well with each other. The uncertainties attributed to the values are quite large due to the error propagation based on the difference of subshell photoionization cross-section ratios. For the CK factor $f_{1,3}$ the relative error of the two other CK factors needs to be included as well, resulting in uncertainties of the order of the extracted value itself. Indeed, these uncertainty budgets are established based on the uncertainties connected to the different factors entering into Eqs. \ref{eq:f23}, \ref{eq:f12} and \ref{eq:f13}.

However, the agreement with other reported CK factors for Gd is not always given. Especially, the comparison with the other reported synchrotron radiation based measurements \cite{Menesguen20} shows a pronounced discrepancy in the sense that especially the CK factors $f_{2,3}$ and $f_{1,2}$ are less than half of the values deduced in the present work. One major aspect to explain this difference is related to the subshell photoionization cross-sections. Indeed, extrapolation of the cross-sections connected to the L$_3$ or L$_2$ beyond the ionization thresholds of the L$_2$ and, respectively the L$_1$ subshells is required. Differences in the data ranges and polynomials used can result in major deviations in the final results. Hence, a critical revision and intercomparison of such aspects as well as of the spectral modelling, semi-empirically versus response function-based, is required in future to assess the origin of such discrepancies.

With regard to other experimentally established values \cite{Papp_1998,Douglas_1972,Gnade_1981}, a better agreement is found for the CK factors $f_{1,2}$ and $f_{1,3}$, where an overlap within the uncertainties of the respective values can be observed, but not necessarily for $f_{2,3}$, where a major inconsistency between the two coincidence measurements was observed and discussed in ref. \cite{Gnade_1981}. With respect to calculations \cite{S.Puri1993} and databases \cite{xraylib,Krause1979}, an agreement within the margin of the uncertainties for the CK factors $f_{1,2}$ and $f_{1,3}$ can be observed. For $f_{2,3}$ the agreement can only be achieved within the combined uncertainty of the presented value and the database value. For databases the consistency of the data used is a measure of the reliability assumed and the uncertainty assigned. For Gd it is for example recommended to use a relative error of 10$\%$ to 20 $\%$ \cite{Krause1979}.

From the reference-free XRF data, the fluorescence yield $\omega_{\text{L}_i}$ of each subshell could also be obtained using the photoionization cross-section $\tau_{\text{L}_i}$ and the deconvolution result $I_{\text{L}_i}$ for the ensemble of emission lines appertaining to an electronic transition to subshell L$_i$
\begin{equation}
    \omega_{\text{L}_i}=\frac{N_{\text{L}_i}(E_0)\sin(\theta) M(E_0,E_{\text{L}_i})}{I_0 \Omega/(4\pi) \epsilon(E_{\text{L}_i}) \tau_{\text{L}_i}(E_0) \rho d}
\end{equation}
for i = 1, 2 or 3 with incident photon intensity $I_0$ and energy $E_0$. For the determination of the fluorescence yield $\omega_{\text{L}_i}$ only the incident photon energy where no CK factors contribute shall be used for the different subshells. The product $\tau_{\text{L}_i}(E_0) \rho d$ was established from the transmission measurements. As during the determination of the CK factors, each subshell fluorescence yield $ \omega_i$ is determined for multiple incident photon energies such that a consistency check can be conducted. For each factor the standard deviation of the results obtained is below 1$\%$. Note that for both, the fluorescence yield and the CK factor, the incident photon energy shall not include energies close to the ionization threshold due to the fine structure on the photoionization cross-section. The results obtained are summarized and compared to literature values in Table \ref{tab:fluoyields}.

\begin{table}[htbp]
  \centering
  \caption{Experimentally determined Gd L subshell fluorescence yields in comparison to available literature sources.}
    \begin{tabular}{|c|c|c|c|}
    \hline
     & $\omega_{\text{L}_3}$ & $\omega_{\text{L}_2}$ & $\omega_{\text{L}_1}$ \\
    \hline
    Present work (RF-XRF) & 0.159(7) & 0.164(11) & 0.076(7) \\
    \hline
    M\'{e}nesguen et al. (2020) \cite{Menesguen20} & 0.159(3) & 0.162(4) & 0.099(3) \\
    \hline
    Puri et al. (1993) \cite{S.Puri1993} & 0.167 & 0.175 & 0.083 \\
    \hline
    xraylib (2012) \cite{xraylib} & 0.155 & 0.158 & 0.102 \\
    \hline
    Krause (1979) \cite{Krause1979} & 0.155 & 0.158 & 0.079 \\
\hline
    Sahnoune et al. (2016) \cite{Y.Sahnoune2016} & 0.162 & 0.1686 & 0.085\\
\hline
    Krishnananda et al. (2016) \cite{Krishnananda} & 0.167(7) & 0.176(8) & 0.089(4) \\
\hline
    Kumar et al. (2010) \cite{Kumar_2010} &  &0.165(13) & 0.101(9) \\
\hline
    Papp et al. (1998) \cite{Papp_1998}  &        &       & 0.101(5) \\
\hline
    Douglas (1972) \cite{Douglas_1972} & 0.187(6) & 0.182(8) &  \\
\hline
    Gnade et al. (1981) \cite{Gnade_1981} & 0.161(19) & 0.159(22)  &  \\
\hline
    \end{tabular}%
  \label{tab:fluoyields}%
\end{table}%

The uncertainty for the fluorescence yields is considerably lower than the one for the CK factors. In contrast to the CK factors a good agreement is obtained with the values from different synchrotron radiation experiments \cite{Menesguen20,Krishnananda}. For $\omega_{\text{L}_3}$ and $\omega_{\text{L}_2}$ the comparisons to most other experimental values \cite{Gnade_1981,Papp_1998,Kumar_2010}, theory \cite{S.Puri1993} and compilations \cite{xraylib,Krause1979,Y.Sahnoune2016} match within the error bars with each other. In ref. \cite{Douglas_1972} larger values are reported, which was already noted in ref. \cite{Gnade_1981} where the same type of experiment was performed. The discrepancies were attributed to the detectors used, background contributions and the data treatment performed \cite{Gnade_1981}. For $\omega_{\text{L}_1}$, most of the different values reported do not agree with and are larger than our result. With respect to experimental works the reason can be the subshell photoionization cross-section $\tau_{\text{L}_1}(E_0)$ \cite{Honicke2016} or the XRF fluorescence rate originating from electronic transitions to the L$_1$ subshell extracted from the experimental data. Without exact information on the these aspects, the situation can currently not be assessed in a more accurate manner.

\section{Conclusions}
The CK factors were determined with a reliable uncertainty using monochromatized synchrotron radiation at different photon energies and two different detection schemes, firstly using an energy-dispersive detector and secondly using a wavelength-dispersive detector. In addition, the fluorescence yields are reported on the basis of reference-free XRF analysis. The uncertainty budget is established relying on experimental values established within this work and only require little input from tabulated data, i.e., the cross-sections for inelastic and elastic scattering with very minor contributions to the uncertainty budget. The agreement between the results and the similarity in relative errors show that for the element investigated both setups are equally well suited. Note that the probability of a radiationless transition was investigated indirectly by means of emitted radiation from the sample, which implies also the quite large errors. CK factors are, indeed, intrinsically difficult to derive due to the required accurate knowledge on the subshell photoionizaition cross-section. The equivalency between the results is relevant since the Coster-Kronig factors are a good probe for theory based on the Hartree-Fock method and the requirement to include relativistic effects \cite{Crasemann84}. Note, that theory applies to single atoms, whereas experiments may be affected by considerable solid-state effects.

A wavelength-dispersive detection scheme, which is more complex to operate compared to the use of an energy-dispersive detector, can be beneficial in different aspects. When turning towards the more complex M or even N-shell spectra a correct assignment of the different emission lines becomes more challenging due to the different transitions involved \cite{McGuire72,Chauchan2008}. Wavelength-dispersive setups have demonstrated their suitability for the accurate determination of the line energies, line widths and relative transition probabilities. In addition, the sensitivity to the chemical state of an element of XES allows to monitor changes in line position or relative transition probabilities, for example due to exchange interactions and crystal effects. Investigations on the dependence of CK factors or fluorescence yields on the chemical state would, however, require improving first the accuracy of the results obtainable. The information on the spectral distribution of the emission lines is required for a correct deconvolution of the experimental spectra and discrimination of the emission lines originating from electronic transitions to different subshells. In addition, the assessment of possible secondary excitation due to XRF appertaining to an electron transition to the inner subshells of an electronic level is facilitated. Such complementary investigations are not always readily obtainable with energy-dispersive detectors, which could therefore profit from such complementary investigations. The strength of energy-dispersive detection schemes is that absolute measurements on the emitted fluorescence intensities are enabled. The latter type of experiment is mandatory for the determination of fluorescence yields and cannot be easily conducted using wavelength-dispersive detection setups, as has been pointed out for the full-cylinder von Hamos geometry used in this study for which the effective solid angle defined as the product of the CCD detection efficiency, the integral reflectivity of HAPG and the solid angle of acceptance of the optic needs to be evaluated in this respect \cite{Malzer2018}. Finally, a correct assessment of the many atomic FP requires even further instrumentation development and the complementary use of energy- and wavelength dispersive detection setups.

\section{Acknowledgements}
Parts of this research was performed within the EMPIR project 19ENV08 AEROMET II. This project has received funding from the EMPIR programme co-financed by the Participating States and from the European Union’s Horizon 2020 research and innovation programme. This project has received funding from the ECSEL Joint Undertaking (JU) IT2 under grant agreement No 875999. The JU receives support from the European Union’s Horizon 2020 research and innovation programme and Netherlands, Belgium, Germany, France, Austria, Hungary, United Kingdom, Romania and Israel.

\bibliography{references}  


\end{document}